\documentclass[journal]{IEEEtran}

\usepackage{amsmath,amssymb}
\usepackage{graphicx}
\usepackage{booktabs}
\usepackage{multirow}
\usepackage{url}
\usepackage[table]{xcolor}
\usepackage{enumitem}
\usepackage{algorithm}
\usepackage{algorithmic}
\usepackage{listings}
\usepackage[most]{tcolorbox}
\tcbuselibrary{listings}
\usepackage[colorlinks=true,linkcolor=blue,citecolor=blue,urlcolor=blue]{hyperref}

\definecolor{jsonkey}{RGB}{24,78,140}
\definecolor{jsonvalue}{RGB}{35,112,76}
\definecolor{jsonnumber}{RGB}{155,55,65}
\definecolor{jsonbg}{RGB}{250,251,252}
\definecolor{jsonhead}{RGB}{235,240,244}
\definecolor{jsonframe}{RGB}{140,150,158}
\definecolor{tablerowgray}{gray}{0.93}

\newcommand{\jkey}[1]{%
  {\color{jsonkey}\bfseries\ttfamily
  \detokenize{"#1"}}}

\newcommand{\jnum}[1]{%
  {\color{jsonnumber}\ttfamily #1}}

\lstdefinelanguage{json}{
    showstringspaces=false,
    breaklines=true,
    keepspaces=true,
    columns=fullflexible,
    string=[s]{"}{"},
    stringstyle=\color{jsonvalue},
    keywords={true,false,null},
    keywordstyle=\bfseries
}

\newtcblisting{jsonprofile}{
    enhanced,
    listing only,
    listing engine=listings,
    colback=jsonbg,
    colframe=jsonframe,
    colbacktitle=jsonhead,
    coltitle=black,
    boxrule=0.55pt,
    arc=1.5mm,
    title={\texttt{JSON}\quad Key-clue Extractor Output},
    fonttitle=\bfseries\footnotesize,
    titlerule=0.4pt,
    left=1.2mm,
    right=1.2mm,
    top=0.8mm,
    bottom=0.8mm,
    listing options={
        language=json,
        basicstyle=\ttfamily\fontsize{6.9}{8.2}\selectfont,
        stringstyle=\color{jsonvalue},
        escapeinside={(*@}{@*)},
        showstringspaces=false,
        breaklines=true,
        keepspaces=true,
        columns=fullflexible
    }
}

\newtcolorbox{definitionbox}{
    enhanced,
    colback=gray!6,
    colframe=gray!45,
    boxrule=0.35pt,
    arc=0.5mm,
    left=1.2mm,
    right=1.2mm,
    top=0.8mm,
    bottom=0.8mm,
    before skip=0.8\baselineskip,
    after skip=0.8\baselineskip
}

\newcommand{\placeholder}[1]{\textcolor{black}{#1}}

\newcommand{\blackcircled}[1]{%
  \tikz[baseline=(C.base)]\node[
    shape=circle,
    fill=black,
    text=white,
    inner sep=0pt,
    minimum size=1.15em,
    font=\scriptsize\bfseries
  ] (C) {#1};%
}

\newtcolorbox{insightbox}[1][]{
  enhanced,
  breakable,                
  colback=gray!10,          
  colframe=gray!60,         
  boxrule=0pt,              
  leftrule=2pt,             
  sharp corners,            
  boxsep=0pt,               
  top=5pt,                  
  bottom=5pt,               
  left=5pt,                
  right=5pt,                
  fontupper=\normalsize,    
  before skip=6pt,         
  after skip=6pt,          
  #1
}

\begin{document}


\title{\textsc{RiskTagger}: Evidence-Guided LLM Agent for Post-Incident Forensic Analysis of Money Laundering in Web3}

\author{Dan Lin,~\IEEEmembership{Member,~IEEE},
        Yanli Ding,
        Weipeng Zou,
        Jiajing Wu,~\IEEEmembership{Senior Member,~IEEE},
        Zhiyin Wu,\\
        Jiachi Chen,
        Xiapu Luo,
        Zibin Zheng,~\IEEEmembership{Fellow,~IEEE}%
\thanks{Manuscript received \placeholder{July xx, 2026}; revised
\placeholder{xxxx}; accepted \placeholder{xxxx}.
This work is supported in part by the
National Natural Science Foundation of China under Grant 62502548, Grant 62372485, and Grant 62332004; in part by the Open Research Fund of The State Key Laboratory of Blockchain and Data Security, Zhejiang University; in part by the Hong Kong RGC Project under Grant PolyU15231223; and in part by  Hong Kong RGC Grant for Theme-based Research Scheme Project under Grant T41-517/25-N. 
\textit{(Corresponding authors: Jiajing Wu and Zhiying Wu)}}%
\thanks{Dan Lin, Yanli Ding, Weipeng Zou, Jiajing Wu, Zhiying Wu, and Zibin
Zheng are with the School of Software Engineering, Sun Yat-sen University,
Zhuhai 519082, China, and the Guangdong Engineering Technology Research
Center of Blockchain, Zhuhai, China \textit{(Email: wujiajing@mail.sysu.edu.cn)}}%
\thanks{Jiachi Chen is with The State Key Laboratory of Blockchain and Data Security, Zhejiang University, Hangzhou, China, and also with the Hangzhou High-Tech Zone (Binjiang) Institute of Blockchain and Data Security, Hangzhou, China.}%
\thanks{Xiapu Luo is with the Department of Computing, The Hong Kong
Polytechnic University, Hong Kong, China.}%
}

\markboth{IEEE Transactions on Information Forensics and Security,~Vol.~xx,
No.~x, xxx~2026}%
{Lin \MakeLowercase{\textit{et al.}}: RiskTagger}

\maketitle

\begin{abstract}
Cryptocurrency money-laundering forensic analysis after Web3 incidents faces challenges such as fragmented evidence, expanding transaction paths, and cross-chain discontinuity. Existing Web3 anti-money-laundering (AML) methods largely rely on manual clues and heuristic or graph-search-based tracing, with outputs typically limited to lists of suspicious addresses and lacking path-level evidence and verifiable explanations. Directly applying general-purpose large language models to raw transaction flows also struggles to ensure evidence constraints and result verifiability.
To address these limitations, this paper presents \textsc{RiskTagger}, an LLM-guided agent for forensic tracing of Web3 cryptocurrency money laundering. \textsc{RiskTagger} embeds the LLM as an evidence-constrained decision component within a controlled tracing loop. It extracts case clues from public incident materials, recursively expands a risk-labeled fund-flow graph over on-chain evidence, and generates evidence-organized reports for analyst review. We evaluate the system on five real-world security incidents spanning multiple years and covering heterogeneous attack patterns and laundering path structures. We further conduct cross-case generalization analysis, baseline comparison, component ablation, and LLM backend analysis.
In the main Bybit case, the system achieves a 97.33\% address recall and a 98.69\% expert-reviewed sampled address precision. Across the other four incidents, it achieves 95.24--100.00\% address recall and 91.27--100.00\% expert-reviewed address precision. The cross-case results further show that the complexity of Web3 money laundering arises from heterogeneous mechanisms, including short-cycle fund fragmentation, long-range laundering paths, interwoven DeFi services, and deterministic denomination splitting. \textsc{RiskTagger} can recover case-related fund paths, identify high-priority risk accounts, and organize public evidence into verifiable forensic reports.

\end{abstract}


\section{Introduction} \label{sec:introduction}

\IEEEPARstart{T}{he} rapid development of Web3 has promoted decentralized finance and cross-chain asset flows~\cite{10273397,lin2025Connector}, but has also made large-scale cryptocurrency theft and subsequent money laundering more difficult to investigate~\cite{Almeida2023ARO}. Web3 money laundering refers to a series of behaviors that conceal the illicit origin of assets and weaken their association with original security incidents~\cite{10.1108/JMLC-02-2020-0018}. For attackers, asset theft is only the first step. The subsequent goal is to hide the fund source and obstruct tracing.

Due to the pseudonymity of public-chain users and the cross-chain mobility of assets, post-incident forensics directly affects asset recovery and subsequent risk control. Therefore, the core problem is: after an incident occurs, how can we rapidly and automatically reconstruct case-related high-risk fund flows and accounts from public information, while providing auditor-friendly explanations and forensic suggestions?


\begin{figure}[t]
    \setlength{\abovecaptionskip}{0.0cm}
    \centering
    \includegraphics[width=0.9\linewidth]{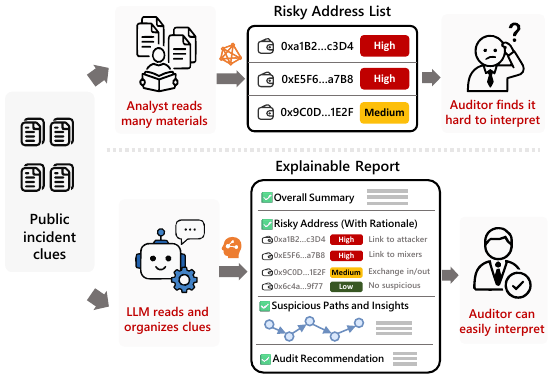}
\caption{Motivation. Manual analysis and existing tracing algorithms often stop at risky-address lists, while auditors require evidence-grounded rationales and transaction-path context.}
    \label{fig:motivation}
\end{figure}

Existing studies have not fully addressed intelligent post-incident forensic analysis. One line of research builds AML datasets or learns graph-based detectors from labeled transaction data~\cite{10.1145/3383455.3422549,Elliptic,Elliptic++,wu2023toward}. These studies are important for detection and benchmarking, but they usually relies on historical labels or predefined transaction features. Another line of work traces or mines illicit account groups using heuristics, graph algorithms, or suspiciousness measures~\cite{Alarab2020,wu2023toward,lin2024denseflow,Xiang2022BABDAB}. These methods can provide useful on-chain signals, but new incidents still require analysts to extract case clues from public materials and organize tracing results into reviewable evidence, which remains labor-intensive. Moreover, public labels are incomplete and may lag behind recent attacks, while older datasets may not cover cross-chain laundering behaviors~\cite{wu2025safeguarding}. Fig.~\ref{fig:motivation} illustrates the motivation of this work.

Large language models (LLMs) create a new opportunity for this setting. Their ability to read unstructured reports, follow context, and produce natural language explanations suggests that they can support incident-driven analysis~\cite{shen2024llmtoolssurvey,li2024surveyllmagent}. However, directly applying an LLM to Web3 laundering tracing is insufficient, and we still face several unique challenges (C):
\begin{itemize}[leftmargin=*]
\item \textbf{C1: High cost and inefficiency of key-clue extraction.} Initial attack addresses, affected assets, abnormal transactions, and timelines are often scattered across long reports, announcements, tables, and explorer pages.
\item \textbf{C2: Complexity of laundering trace inference.} laundering paths are multi-step and multi-chain, so unconstrained language reasoning over raw transaction streams can hallucinate or make inconsistent decisions when it lacks on-chain behavioral constraints.
\item \textbf{C3: Lack of interpretable results.} A list of suspicious addresses alone is not enough for forensic use. Analysts need paths, risk rationales, and evidence anchors that explain why an account or transaction remains relevant to the case.
\end{itemize}

To address these challenges, we design \textsc{RiskTagger}, an LLM-guided agent for Web3 post-incident laundering forensic from public evidence. \textsc{RiskTagger} is organized around three coordinated components. 
The \textit{Key-clue Extractor} first automatically recovers structured starting information from public incident materials, including seed addresses, affected assets, and initial event context. 
The \textit{Laundering Tracer} then iteratively fetches public on-chain records, translates transaction evidence, reasons about account risk, and determines the next graph frontier. Unlike treating an LLM as a standalone classifier, \textsc{RiskTagger} couples risk decisions with recursive graph expansion: the model's judgment affects which accounts are examined next, while the \textit{expansion controller} bounds this process. 
Finally, the \textit{Evidence Explainer} organizes the traced results into a case report, including a forensic summary, risk-account records, suspicious transaction behaviors, and evidence-grounded forensic suggestions for analyst review.
This paper makes the following contributions:

\begin{itemize}[leftmargin=*]
\item \textbf{AML Forensics.} We define an evidence-driven Web3 post-incident AML forensic task: given public incident materials, confirmed seed addresses, and public on-chain data, the system should recover case-related fund flows, prioritize risk accounts, and produce reviewable evidence.
\item \textbf{Guided Architecture.} We propose \textsc{RiskTagger}\footnote{Available at \url{https://github.com/Connector-Tool/RiskTagger}}, a closed-loop LLM-guided framework that turns incident clues into recursive graph analysis, with evidence translation, constrained risk reasoning, controlled expansion, and evidence-organized reporting.
\item \textbf{Empirical Evaluation.} We evaluate \textsc{RiskTagger} on five real-world security incidents with reported losses ranging from \$11.6M to \$1.5B, totaling about \$2.39B. The evaluation covers main-case analysis, cross-case generalization, tracing-baseline comparison, component ablation, and LLM backend comparison.
\item \textbf{Forensic Insights.} Beyond recovering accounts and paths, \textsc{RiskTagger} surfaces reviewable laundering patterns. For example, in the Bybit case, it identifies a bimodal pulsing topology: early native-token movement is followed by later stablecoin-oriented fragmentation through DeFi services.
\end{itemize}


\section{Background} \label{sec:background}

\noindent\textbf{Web3 transaction basics.} In Web3, every transfer or contract interaction is recorded as a public, immutable transaction on a blockchain. Each transaction includes metadata such as a unique hash, timestamp, sender and receiver addresses, transferred tokens, and related event logs~\cite{Wu2023MoTs}. Smart contracts~\cite{ZHENG2020475} are self-executing programs deployed on-chain, enabling programmable financial operations such as token swaps or cross-chain transfers~\cite{lin2025Connector}. Although these records are verifiable and permanently stored, user identities remain pseudonymous, and relationships between addresses must be inferred from transaction patterns rather than explicit ownership~\cite{10.1108/JFC-06-2020-0113}. This transparency-anonymity trade-off forms the foundation and difficulty of blockchain-based AML research.

\noindent\textbf{AML paradigm and red flags.} Anti-money laundering (AML) analysis generally follows a three-stage paradigm: placement, layering, and integration~\cite{PLI}. In the first stage, illicit funds are introduced into the financial system; during layering, their origins are obscured through multi-chain transfers, token swaps on decentralized exchanges (DEXs)~\cite{dex}, stablecoin conversions, and cross-chain bridge movements; finally, the funds are reintegrated into the economy under seemingly legitimate forms. Regulatory bodies such as the Financial Action Task Force (FATF)~\cite{10.1108/JMLC-06-2020-0070} have established a set of red-flag indicators to identify suspicious behaviors, including unusually large or frequent transactions, interactions with mixers or exploit-linked wallets, and the aggregation or dispersion of assets across multiple chains. These indicators form core cues for detecting Web3 money-laundering activities and serve as reasoning signals within the \textsc{RiskTagger} framework.


\section{Problem Setting and Threat Model} \label{sec:problem}

We study post-incident analysis of cryptocurrency laundering networks from public evidence. Given a disclosed Web3 theft, exploit, or fraud incident, the analyst starts from public incident materials and public on-chain records. The task is to reconstruct case-related fund flows and identify accounts that should be prioritized for forensic review.

\begin{definitionbox}
\noindent\textbf{Problem Definition.} \textit{Let $\mathcal{D}$ denote the public incident materials, $\mathcal{S}_0$ the confirmed seed addresses extracted from the incident materials. The post-incident analysis system produces a case profile $\mathcal{P}$, a risk-labeled tracing graph $\mathcal{G}=(\mathcal{V},\mathcal{E})$, and an evidence-organized report $\mathcal{R}$. Each vertex in $\mathcal{G}$ represents an address, and each edge records a fund-flow relation supported by public evidence.}
\end{definitionbox}

\begin{figure*}[h]
    \centering
    \setlength{\abovecaptionskip}{0.0cm}
    \includegraphics[width=0.9\textwidth]{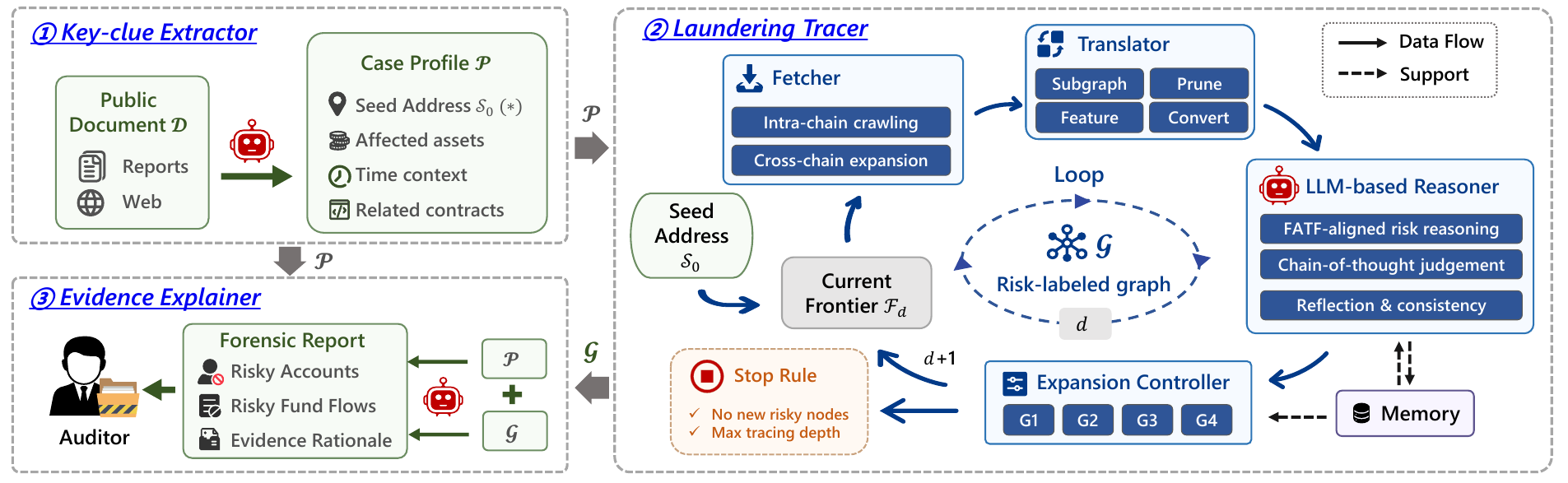}
    \caption{System overview of \textsc{RiskTagger}. Key-clue Extractor builds the case profile, Laundering Tracer performs the closed-loop fund-flow tracing, and Evidence Explainer generates reviewable forensic reports. }
    \label{fig:method_overview}
\end{figure*}


\noindent\textbf{Adversary objective.} We consider a post-incident laundering adversary who uses publicly visible on-chain activity to obscure the provenance of assets obtained from theft, exploits, or fraud. The objective is to weaken incident linkage and increase forensic tracing cost. The adversary may disperse funds across controlled addresses, route assets through DeFi services or cross-chain bridges, and inject low-value noisy transfers to enlarge the tracing space. 

\noindent\textbf{Adversary limitations.} The adversary cannot alter historical blockchain records, break the underlying cryptographic mechanisms, or control the public ledgers. \textsc{RiskTagger} therefore treats public chain records as stable transaction evidence, while recognizing that public labels and incident reports may be incomplete, delayed, or partially inaccurate.

\noindent\textbf{Defender assumption.} The defender is an analyst using offline forensic tools after an incident has been known. The system assists the analyst by organizing public evidence into traceable paths and accounts for subsequent forensic review.


\section{Related Work} 
\label{sec:related_work}

\noindent\textbf{Anti-money laundering on Blockchain.}
Prior blockchain AML studies mainly focus on suspicious-transaction detection, labeled datasets, and graph-based illicit-account analysis. Alarab \textit{et al.}~\cite{Alarab2020} compared supervised learning methods for Bitcoin AML, and Lorenz \textit{et al.}~\cite{10.1145/3383455.3422549} studied money-laundering detection under label scarcity. The Elliptic dataset and Elliptic++ provide large-scale Bitcoin benchmarks for financial forensics~\cite{Elliptic,Elliptic++}, while BABD studies Bitcoin address behavior patterns~\cite{Xiang2022BABDAB}. Graph-based studies further model illicit activities through transaction graphs, temporal features, or representation learning~\cite{Humranan2023ASO,nicholls2023fraudlens,poursafaei2021sigtran}. Recent TIFS studies extend this direction by learning cross-platform interaction features, dynamic Ethereum account representations, and generative or contrastive self-supervision for malicious-account or fraud detection~\cite{10979464,11173689,10812703}. Related TDSC studies further address Ethereum phishing detection with transaction subgraphs, DeFi flash-loan attack detection with hypergraph neural networks, and hidden cryptomining detection~\cite{10938293,11539006,11371732}. For account-level tracing in Ethereum, Wu \textit{et al.}~\cite{wu2023toward} analyzed asset flows in Ethereum heists, Wu \textit{et al.}~\cite{wu2023toward} mined unlabeled laundering groups, and DenseFlow identifies Ethereum laundering groups through suspiciousness metrics and flow analysis~\cite{lin2024denseflow}. These works provide important datasets and on-chain analysis methods. In contrast, \textsc{RiskTagger} targets post-incident forensic tracing from public incident materials and public on-chain records, emphasizing dynamic recovery of case-related fund paths and evidence that can be reviewed by analysts.

\noindent\textbf{LLMs and agents for blockchain security.}
Large Language Models will demonstrate significant potential in enhancing the fraud detection capabilities of blockchain networks and applications. BERT4ETH \cite{BERT4ETH} is a pre-trained Transformer model specifically designed for Ethereum fraud detection tasks. Meanwhile, LLMs possess strong text comprehension and cross-semantic reasoning capabilities, endowing them with distinct advantages in the field of smart contract analysis. For instance, Luo \textit{et al.}~ \cite{luo2024fellmvp} classified smart contract vulnerabilities by integrating multiple LLM agents. Liu \textit{et al.}~ proposed PropertyGPT\cite{liu2024propertygpt}, which improves the efficiency of formal verification by automatically generating smart contract properties. In the field of anomaly detection,  Yu \textit{et al.}~ \cite{yu2024blockfound} proposed a blockchain-specific multi-modal Tokenizer and a Masked Language Modeling-based foundation model, enabling cross-chain anomaly detection. Gai \textit{et al.}~ \cite{Gai2023BlockchainLL} implemented real-time blockchain anomaly detection by training an LLM from scratch.
\textbf{However, existing LLM-based studies primarily focus on fraud or contract-level analysis, while post-incident laundering forensic remains largely unexplored.}


\section{Method} \label{sec:method}

\subsection{Overview} \label{subsec:method_overview}

The key design of \textsc{RiskTagger} is to use the LLM as an evidence-conditioned decision component rather than an unconstrained generator over raw transaction streams. Each model decision is conditioned on the case profile, fetched transactions, structured account-level evidence, historical memory, and explicit expansion rules. As shown in Fig.~\ref{fig:method_overview}, \textsc{RiskTagger} consists of three core components:
\begin{itemize}[leftmargin=*]
\item \textbf{Key-clue extractor.} Given public incident materials $\mathcal{D}$, the key-clue extractor consolidates seed addresses $\mathcal{S}_0$, affected assets, time context, related contracts, and initial fund-flow clues. It produces a structured case profile $\mathcal{P}$ that provides the initial evidence context for subsequent tracing.
\item \textbf{Laundering tracer.} Given the case profile $\mathcal{P}$, and public on-chain data $\mathcal{X}$, the laundering tracer recursively constructs a risk-labeled graph $\mathcal{G}$. At each tracing depth, it fetches transaction evidence, translates raw records into compact account-level evidence, assigns risk labels through constrained reasoning, and selects the next frontier.
\item \textbf{Evidence explainer.} Given the case profile $\mathcal{P}$ and the risk-labeled graph $\mathcal{G}$, the evidence explainer generates an evidence-organized report $\mathcal{R}$. The report summarizes laundering paths, high-priority accounts, representative transactions, and uncertainty for forensic review.
\end{itemize}

\begin{figure}[t]
    \centering
    \includegraphics[width=\columnwidth]{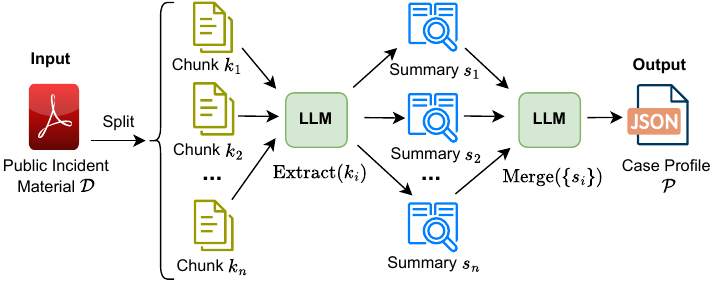}
\caption{Workflow of the Key-clue Extractor. Public incident materials $\mathcal{D}$ are split into chunks, summarized into local clues $s_i$, and merged into the structured case profile $\mathcal{P}$.}
    \label{fig:key_clue_extractor}
\end{figure}

\subsection{Key-clue Extractor} \label{subsec:key_clue_extractor}

Key-clue extractor converts public incident materials $\mathcal{D}$ into a case profile $\mathcal{P}$ by extracting local clues from document chunks and consolidating them into verified case-level evidence for downstream tracing, as shown in Fig.~\ref{fig:key_clue_extractor}.

\begin{itemize}[leftmargin=*]
\item \textbf{Document splitting.} The system segments $\mathcal{D}$ into chunks $\mathcal{K}=\{k_i\}$ by paragraph, page, or semantic boundary, while recording each chunk's source location.

\item \textbf{Chunk-level extraction.} For each chunk $k_i$, the system generates a local summary $s_i=\mathrm{Extract}(k_i)$, covering seed addresses, related contracts, transaction hashes, asset types, amounts, time windows, and initial fund-flow clues.

\item \textbf{Global consolidation.} The system derives the case profile $ \mathcal{P}=\mathrm{Merge}(\{s_i\}_{i=1}^{|\mathcal{K}|}).$ by normalizing entities, removing duplicates, merging cross-chunk references, and resolving conflicts by keeping the better-supported candidate or marking the field as uncertain.
\end{itemize}

\begin{figure}[t]
\centering
\begin{minipage}{\columnwidth}
\begin{jsonprofile}
{
  (*@\jkey{project_info}@*): {
    (*@\jkey{event_name}@*): "Bybit Cold Wallet Hack",
    (*@\jkey{date}@*): "2025-02-21"
  },
  (*@\jkey{findings}@*): [{
    (*@\jkey{attack_vector}@*):
      "Supply-chain compromise via malicious
       JavaScript injected into Safe{Wallet}...",
    (*@\jkey{affected_platform}@*):
      "Bybit via Safe{Wallet}",
    (*@\jkey{chain}@*): ["Ethereum"],
    (*@\jkey{contract_address}@*): ["0xbDd0...9516"],
    (*@\jkey{attacker_addresses}@*):
      ["0x0fa0...6d2e", "...", "0xdd90...5f92"],
    (*@\jkey{victim_addresses}@*): ["0x1Db9...FCF4"],
    (*@\jkey{stolen_amount_usd}@*): (*@\jnum{1500000000}@*),
    (*@\jkey{stolen_amount_token}@*):
      {"ETH": (*@\jnum{401000}@*), "mETH": (*@\jnum{8000}@*), "...": "..."},
    (*@\jkey{evidence_snippets}@*):
      ["Bybit wallet 0x1Db9...FCF4 was
        compromised on 2025-02-21.",
       "...",
       "Safe infrastructure was compromised
        through the malicious script app-...js."]
  }]
}
\end{jsonprofile}
\end{minipage}
\caption{Abbreviated structured output generated by the Key-clue Extractor for the Bybit incident. Long addresses, repeated entries, and intermediate evidence are shortened for readability.}
\label{fig:extractor_json}
\end{figure}


The final output is a structured JSON case profile, as illustrated in Fig.~\ref{fig:extractor_json}. The profile does not fill unsupported fields, and it retains sources or evidence snippets for key clues to support downstream tracing and human review.

\subsection{Laundering Tracer} \label{subsec:laundering_tracer}

The laundering tracer reconstructs case-related fund flows from the case profile and public on-chain data. We use tracing depth to denote the recursive expansion layer from the seed frontier; this depth is analogous to a hop in graph search. Given $\mathcal{P}$ and $\mathcal{S}_0$, the tracer maintains a risk-labeled graph $\mathcal{G}=(\mathcal{V},\mathcal{E})$ and a tracing frontier $\mathcal{F}_d$ at current tracing depth $d$. For each address in the current frontier, the tracer executes four stages, namely the transaction fetcher, transaction translator, memory-augmented LLM reasoner, and expansion controller, as organized in Fig.~\ref{fig:method_overview}. The loop stops when the frontier is empty, the configured tracing budget is reached, or no new case-relevant accounts are found. Algorithm~\ref{alg:tracer} summarizes this recursive tracing loop.

\subsubsection{Fetcher} \label{subsubsec:transaction_fetcher}

The transaction fetcher constructs an evidence-preserving local transaction view for each address in the current candidate set. It is not a risk classifier. Its role is to collect verifiable on-chain records and expose the transaction-level basis used by the translator, reasoner, and expansion controller. 

\textbf{(1) Intra-chain crawling.} For each candidate address obtained from the case profile or previous tracing depths, \textsc{RiskTagger} retrieves intra-chain transaction data using \textsf{BlockchainSpider}\footnote{\url{https://github.com/wuzhy1ng/BlockchainSpider}}. \textsf{BlockchainSpider} is a Scrapy-based blockchain data crawler that supports address-centered collection from EVM-compatible chains such as Ethereum, BSC, and Polygon. It fetches upstream and downstream transactions associated with the given address. The output is a local transaction subgraph, whose nodes represent accounts or contracts and whose edges represent transfers or function calls. In addition, the fetcher preserves a canonical transaction record table with fields including \texttt{hash}, \texttt{from}, \texttt{to}, \texttt{value}, \texttt{timeStamp}, \texttt{blockNumber}, \texttt{tokenSymbol}, and \texttt{contractAddress}, together with auxiliary fields such as \texttt{isError}, \texttt{input}, \texttt{nonce}, \texttt{blockHash}, \texttt{gas}~\cite{zarir2021developing}, \texttt{gasPrice}, \texttt{gasUsed}, and \texttt{confirmations}.

\textbf{(2) Cross-chain expansion.} When a candidate's outgoing flow enters a bridge-related contract, \textsc{RiskTagger} adopts a two-level association strategy. If the bridge provides a public API wrapper that can return the destination-chain result from a source-chain deposit transaction, the system first uses this fast path. For broader coverage, \textsc{RiskTagger} can also invoke \textsc{Connector}\footnote{\url{https://github.com/Connector-Tool/Connector}}~\cite{lin2025Connector}. \textsc{Connector} identifies deposit transactions from public on-chain traces, event logs, transfer patterns, and contract execution traces, and matches them with corresponding withdrawals across chains. The output is a labeled bridge-level edge $(tx_{\mathrm{src}}, tx_{\mathrm{dst}})$ enriched with metadata including amount, token, timestamp, source chain, destination chain, and destination address. If neither path provides public evidence, the fetcher records the boundary instead of inferring an unsupported continuation.

\begin{algorithm}[t]
\footnotesize
\caption{Evidence-Guided Laundering Tracer}
\label{alg:tracer}
\begin{algorithmic}[1]
\REQUIRE Case profile $\mathcal{P}$, seed addresses $\mathcal{S}_0$, on-chain data $\mathcal{X}$, maximum tracing depth $D$
\ENSURE Risk-labeled tracing graph $\mathcal{G}$
\STATE $\mathcal{G}\leftarrow(\mathcal{S}_0,\emptyset)$; $\mathcal{F}_0\leftarrow\mathcal{S}_0$; $\mathcal{V}_{seen}\leftarrow\emptyset$; $d\leftarrow0$
\WHILE{$d < D$ and $\mathcal{F}_d \neq \emptyset$ }
    \STATE $\mathcal{F}_{d+1}\leftarrow\emptyset$
    \FORALL{$a\in\mathcal{F}_d\setminus\mathcal{V}_{seen}$}
        \STATE $\mathcal{V}_{seen}\leftarrow\mathcal{V}_{seen}\cup\{a\}$
        \STATE $\mathcal{T}_a\leftarrow\textsc{Fetch}(a,\mathcal{P},\mathcal{X})$
        \STATE $\mathcal{J}_a\leftarrow\textsc{TranslateToJSON}(\mathcal{T}_a,\kappa)$
        \STATE $\mathcal{M}_a\leftarrow\textsc{RetrieveMemory}(a,\mathcal{J}_a,\mathcal{G})$
        \STATE $(\ell_a,\gamma_a,b_a)\leftarrow\textsc{ReasonAndReflect}(\mathcal{J}_a,\mathcal{M}_a,\mathcal{P})$
        \STATE $\mathcal{G}\leftarrow\textsc{UpdateGraph}(\mathcal{G},a,\mathcal{T}_a,\ell_a,\gamma_a)$
        \STATE $\textsc{UpdateMemory}(a,\mathcal{J}_a,\ell_a,\gamma_a)$
        \IF{$b_a=\mathrm{true}$}
            \STATE $\mathcal{C}_a\leftarrow\textsc{ControlExpansion}(\mathcal{T}_a,\mathcal{J}_a,\ell_a,\mathcal{P},\mathcal{V}_{seen})$
            \STATE $\mathcal{F}_{d+1}\leftarrow\mathcal{F}_{d+1}\cup\mathcal{C}_a$
        \ENDIF
    \ENDFOR
    
    $d \leftarrow d+1 $
\ENDWHILE
\RETURN $\mathcal{G}$
\end{algorithmic}
\end{algorithm}

\subsubsection{Translator} \label{subsubsec:transaction_translator}

Raw transaction records collected by the fetcher are heterogeneous and too verbose for stable LLM reasoning. The transaction translator converts the local transaction subgraph $\mathcal{G}_a$ of each candidate address $a$ into a compact account-level evidence profile $\mathcal{J}_a=\textsc{TranslateToJSON}(\mathcal{G}_a,\kappa)$. This profile is the structured evidence interface between on-chain records and the reasoner.

\textbf{(1) Subgraph construction and pruning.} For each candidate address, the translator builds a local transaction subgraph centered on $a$. To control context length, transactions are ranked by value salience, temporal recency, upstream risk relation, token type, and cross-chain continuity. The translator then retains up to $\kappa$ representative records while preserving links to their original transaction hashes.

\textbf{(2) Feature embedding and format conversion.} The translator extracts account-level behavioral features from the pruned subgraph, including incoming and outgoing transaction counts, total transferred value, dominant assets, counterparty diversity, fan-out ratio, holding time, transfer frequency, and bridge-related events. It also injects available label context, such as public labels, expert-provided labels, and \textsc{RiskTagger}'s historical AI labels, as evidence rather than deterministic rules. The final output is a \texttt{JSON}-style account profile containing representative transactions, behavioral features, upstream/downstream relations, and label context.

\subsubsection{Memory-Augmented LLM Reasoner} \label{subsubsec:reasoner}

The memory-augmented LLM reasoner assigns a case-specific risk label and an expansion recommendation to each candidate account. Given the translated account profile $\mathcal{J}_a$, retrieved memory $\mathcal{M}_a$, and case profile $\mathcal{P}$, the reasoner computes $(\ell_a,\gamma_a,b_a)=\textsc{ReasonAndReflect}(\mathcal{J}_a,\mathcal{M}_a,\mathcal{P})$, where $\ell_a$ is the risk label, $\gamma_a$ is the evidence-grounded rationale, and $b_a$ indicates whether downstream expansion should continue.

\textbf{(1) Memory-augmented evidence retrieval.} Before reasoning, \textsc{RiskTagger} retrieves compact historical memory for the current account or similar accounts observed at prior tracing depths. The memory stores previous risk labels, salient evidence dimensions, and graph context, rather than full raw transaction logs. This keeps the reasoner consistent across repeated infrastructure, recurring laundering patterns, and addresses revisited through different paths.

\textbf{(2) CoT-constrained risk reasoning.} The reasoner follows a constrained Chain-of-thought (CoT)-style decision process~\cite{sprague2025cotcotchainofthoughthelps}, guided by FATF virtual-asset red-flag indicators~\cite{fatf2020}. 
The analysis covers four dimensions:
\begin{itemize}[leftmargin=*]
    \item \textbf{Transaction patterns}: Detection of unusual behaviors, 
    such as high-frequency or fragmented transfers within short time.
    \item \textbf{Fund flows}: Identification of aggregation-dispersion patterns, 
    where funds are pooled from multiple sources and quickly dispersed to many destinations.
    \item \textbf{Associated addresses}: Links to known high-risk entities, 
    including mixers, darknet addresses, or zero-history wallets.
    \item \textbf{Temporal signs}: Abnormal timing (e.g., large transfers at night) 
    or sudden changes in transaction frequency.
\end{itemize}

The output label is selected from $\{\textit{high}, \textit{medium}, \textit{low}, \textit{none}\}$, and the rationale must cite concrete observations. The prompt template is shown in the Supplemental Material.

\textbf{(3) Reflection and consistency check.} To reduce hallucination and label inconsistency, \textsc{RiskTagger} applies a reflection step after the initial decision~\cite{ji2023towards}. The reflection checks whether the assigned risk label conflicts with transaction evidence, retrieved memory, known labels, timing constraints, or benign service behavior. If conflicts remain, the system revises the label, lowers confidence, suppresses expansion, or marks the account as uncertain.


\begin{table}[t]
    \centering
\caption{Output schema of the evidence explainer.}
    \label{tab:explainer_schema}
    \footnotesize
    \rowcolors{2}{tablerowgray}{white}
    \renewcommand{\arraystretch}{1.3}
    \begin{tabular}{
        >{\raggedright\arraybackslash}m{0.28\columnwidth}
        >{\raggedright\arraybackslash}m{0.65\columnwidth}
    }
        \hline
        \textbf{Record type} & \textbf{Evidence-anchored fields} \\
        \hline
        Case summary & Event source, affected assets, seed addresses, time context, and initial flow clues from $\mathcal{P}$. \\
        Risk-account record & Address, risk label, upstream source, downstream behavior, representative transactions, and rationale anchors. \\
        Fund-flow path & Ordered addresses and transactions, bridge-level edges, assets, timestamps, and path boundary markers. \\
        Pattern summary & Repeated graph motifs and supporting examples, such as splitting, aggregation, pass-through, asset conversion, or bridge movement. \\
        Uncertainty log & Missing bridge matches, incomplete paths, conflicting labels, or statements without sufficient public evidence. \\
        \hline
    \end{tabular}
    \rowcolors{2}{}{}
\end{table}

\subsubsection{Expansion Controller} \label{subsec:expansion_controller}

The expansion controller constructs the next tracing frontier $\mathcal{F}_{d+1}$. Its input is the reasoner's label and expansion recommendation for each current address, together with the translated downstream transaction evidence. Its output is a bounded set of candidate addresses $\mathcal{C}_a$ to examine at the next tracing depth. Operationally, the controller is implemented as four sequential evidence-preserving gates that transform a noisy candidate set into a traceable frontier.

\textbf{\blackcircled{G1} Feasibility gate.} 
The first gate keeps only downstream transfers that are feasible under the current case profile. A retained transfer edge $e$ must involve a case-relevant asset, exceed the dust threshold $\tau$, and occur within a local time window after the upstream receipt, i.e., $v(e)\ge\tau$, and $\Delta t(e)\le T$. For transfers involving different tokens, $v(e)$ is normalized using the token price queried from DeFiLlama by the token address and the incident timestamp. 

\textbf{\blackcircled{G2} Semantic routing gate.} The second gate routes each feasible receiver according to its on-chain semantics. \textsc{RiskTagger} uses public Scan APIs and bridge evidence to distinguish externally owned accounts, bridge endpoints, service-like entities, and unrelated contracts. Ordinary accounts are passed to the next gate; bridge interactions are converted into target-chain continuations when public evidence supports the jump; unrelated contracts are treated as boundaries. Service-like entities are retained only when case evidence supports downstream tracing; otherwise, they are bounded or deprioritized for analyst review.

\textbf{\blackcircled{G3} Fund-dominance gate.}
For ordinary downstream expansion, the controller keeps the smallest receiver set $\mathcal{A}_{\mathrm{recv}}'(a) \subseteq \mathcal{A}_{\mathrm{recv}}(a)$ whose cumulative value covers a dominant fraction of the outgoing flow, i.e., $$ \sum_{u\in\mathcal{A}_{\mathrm{recv}}'(a)} v(a,u) / \sum_{u\in\mathcal{A}_{\mathrm{recv}}(a)} v(a,u) \ge \theta.$$
Here, $\mathcal{A}_{\mathrm{recv}}(a)$ is the candidate receiver set, $v(a,u)$ is the aggregated value transferred from account $a$ to receiver $u$, and $\theta$ is the fund-dominance threshold, set to $0.99$ in our evaluation. This gate preserves the main fund path while suppressing dust-level branches and adversarial fan-out noise.


\textbf{\blackcircled{G4} Loop validity gate.} Before adding an address to $\mathcal{F}_{d+1}$, \textsc{RiskTagger} applies an \textsc{IsValid} check. The gate removes previously visited nodes, known legitimate entities when public labels support exclusion, candidates outside the local time window, and edges that would introduce uninformative cycles.

The controller is not an after-the-fact cleanup step. It defines the operational boundary of recursive tracing. By combining the reasoner's account-level judgment with transaction-level constraints, \textsc{RiskTagger} avoids uncontrolled graph expansion while keeping the loop focused on accounts that plausibly carry case-related funds.

\subsection{Evidence Explainer} \label{subsec:evidence_explainer}

The objective of the explainer is to transform the extracted clues and risk assessments from the previous modules into auditor-friendly explanatory documents. Following the output schema in Table~\ref{tab:explainer_schema}, the module first generates a case summary from the extractor output, including the time range, involved chains and assets, and the source of illicit funds. It then integrates the tracer results to highlight high-risk accounts and suspicious transaction paths, while annotating key transaction types such as cross-chain transfers, mixing~\cite{TIPPE2025301876}, stablecoin~\cite{mita2019stablecoin} aggregation, and token swaps~\cite{su14020913}. Finally, the module leverages an LLM to convert structured results into natural-language narratives, summarizing core laundering patterns and ensuring that the annotations are transparent, traceable, and interpretable.
The prompt design is shown in the Supplement Material.
\section{Evaluation Setup} \label{sec:evaluation_setup}

In this section, we evaluate and analyze \textsc{RiskTagger} by answering the following research questions (RQs):
\begin{itemize}[leftmargin=*]
    \item \textbf{RQ1. End-to-end forensic effectiveness on the Bybit Hack.} 
    How well does \textsc{RiskTagger} extract case clues, recover laundering-related accounts, and generate evidence-organized reports in the Bybit Hack (the largest reported cryptocurrency theft to date)~\cite{guardian2025bybitLargestTheft} ?

    \item \textbf{RQ2. Cross-case generalization.} 
    Does \textsc{RiskTagger} generalize beyond the Bybit Hack to heterogeneous laundering incidents with different tracing depths, assets, protocols, and service boundaries?

    \item \textbf{RQ3. Comparison with tracing baselines.} 
    Compared with representative tracing baselines, does \textsc{RiskTagger} provide more reliable tracing outputs in terms of address-level accuracy and fund-flow recovery?

    \item \textbf{RQ4. Component contribution and backend sensitivity.} 
    Which components of \textsc{RiskTagger} contribute to tracing stability, and how sensitive is the system to different LLM backends?
\end{itemize}


Table~\ref{tab:case_overview} summarizes the evaluation cases. We use the Bybit Hack as the primary incident and four earlier incidents to test cross-case behavior under different attack vectors and laundering scales. Publicly labeled case-related accounts, collected from Etherscan label/search pages and screened by keyword, serve as the ground truth (GT) for recall evaluation.


\begin{table}[t]
\centering
\caption{Overview of evaluation cases.}
\label{tab:case_overview}
\scriptsize
\setlength{\tabcolsep}{2.4pt}
\renewcommand{\arraystretch}{1.3}
\rowcolors{2}{tablerowgray}{white}
\begin{tabular}{p{0.25\columnwidth}p{0.31\columnwidth}p{0.13\columnwidth}p{0.20\columnwidth}}
\hline
\textbf{Case} & \textbf{Incident context} & \textbf{Loss} &
\textbf{GT} \\
\hline
Bybit Hack, 2025\footnotemark[1] & Contract-logic hijack &
$\sim$\$1.5B & 73~(\href{https://etherscan.io/accounts/label/bybit-exploit}{Etherscan}) \\
Li.Fi, 2024\footnotemark[2] & DeFi smart-contract exploit &
$\sim$\$11.6M & 17~(\href{https://etherscan.io/search?q=Li.Fi\%20exploit\&filter=address}{Etherscan}) \\
HTX \& Heco, 2023\footnotemark[3] & Cross-chain bridge attack &
$\sim$\$110M & 21~(\href{https://etherscan.io/search?q=Heco\%20Bridge\%20Exploiter\&filter=address}{Heco},
\href{https://etherscan.io/search?q=HTX\%20Exploiter\&filter=address}{HTX}) \\
Ronin, 2022\footnotemark[4] & Bridge key compromise &
$\sim$\$624M & 9~(\href{https://etherscan.io/search?q=Ronin+exploit\&filter=address}{Etherscan}) \\
Vulcan Forged, 2021\footnotemark[5] & Platform intrusion &
$\sim$\$140M & 2~(\href{https://etherscan.io/search?q=Vulcan\%20Forged\%20\%20hack\&filter=address}{Etherscan}) \\
\hline
\end{tabular}
\rowcolors{2}{}{}
\vspace{0.5mm}
\begin{minipage}{0.96\columnwidth}
\footnotesize
\end{minipage}
\end{table}
\footnotetext[1]{Rekt, ``Leaderboard''~\href{https://rekt.news/zh/leaderboard}{[Online]}.} \footnotetext[2]{MetaSleuth, ``Illicit Fund Flow Case Study: LiFi Attack''~\href{https://metasleuth.io/blog/illicit-fund-flow-case-study-lifi-attack-metasleuth}{[Online]}.} \footnotetext[3]{Hacken, ``HECO Bridge Hack Explained''~\href{https://hacken.io/insights/heco-bridge-hack-explained/}{[Online]}.} \footnotetext[4]{BitPinas, ``Tracking the Stolen Funds from Ronin Network Using Breadcrumbs''~\href{https://bitpinas.com/feature/tracking-the-stolen-funds-from-ronin-network-using-breadcrumbs/}{[Online]}.} \footnotetext[5]{Smart Contracts Hacking, ``Vulcan Forged Hack 2021''~\href{https://smartcontractshacking.com/hacks/vulcan-forged-hack-2021}{[Online]}.}

\begin{table}[t]
\centering
\caption{Quantitative performance on the Bybit Hack.}
\label{tab:bybit_performance}
\footnotesize
\setlength{\tabcolsep}{3.2pt}
\renewcommand{\arraystretch}{1.3}
\begin{tabular}{lll}
\hline
\textbf{Module} & \textbf{Metric} & \textbf{Result} \\
\hline
\multirow{2}{*}{Extractor} & Entity precision & 100.00\% \\
& Information coverage & 100.00\% \\
\hline
\multirow{8}{*}{Tracer} & Max depth & 15 \\
& $|\text{New}~\mathcal{A}_{+}|$ & 14,952 \\
& Fund recall $R_f$ & 89.07\% \\
& Address recall $R_a$ & 97.33\% \\
& Address precision $P_a$ & 98.69\% \\
& False positive rate & 14.71\% \\
& F1-score & 98.01\% \\
& Krippendorff's $\alpha$ & 0.7848 \\
\hline
\multirow{5}{*}{Explainer} & Accuracy & 4.00/5 \\
& Logic & 4.86/5 \\
& Professionalism & 4.86/5 \\
& Actionability & 4.86/5 \\
& Overall & 4.57/5 \\
\hline
\end{tabular}
\vspace{0.5mm}
\begin{minipage}{0.96\columnwidth}
\footnotesize
\end{minipage}
\end{table}

\begin{table}[t]
\setlength{\belowcaptionskip}{0pt}   
  \centering
\caption{Information extraction performance of \textsc{RiskTagger}.}
 \label{tab:rq1-case-example}
  \scalebox{0.8}{
  \renewcommand\arraystretch{1.2} 
      \begin{tabular}{p{2cm} p{3cm} p{3cm} p{1.2cm}}
        \toprule
        \textbf{Entity} & \textbf{Meaning} & \textbf{\textsc{RiskTagger} Result} & \textbf{Experts} \\
        \midrule
        \rowcolor{gray!10} 
        \textit{chain} & Affected blockchain network  & Ethereum & \checkmark{} / \checkmark{}   \\
        \textit{attack\_vector} & Attack techniques and exploited vulnerabilities &  Supply chain compromise via malicious JavaScript injection in Safe\{Wallet\} frontend. DELEGATECALL-based contract logic hijacking. & \checkmark{} / \checkmark{} \\
        \rowcolor{gray!10} 
        \textit{affected\_platform} & Impacted platform or service (exploited or affected entity) & Bybit (via compromised Safe\{Wallet\} infrastructure) & \checkmark{} / \checkmark{} \\
        \textit{contract\_address} & Involved or exploited smart contract addresses  
        & \href{https://etherscan.io/address/0xbDd077f651EBe7f7b3cE16fe5F2b025BE2969516}{\textit{0xbDd0}}  ,\href{https://etherscan.io/address/0x96221423681A6d52E184D440a8eFCEbB105C7242}{\textit{0x9622}} & \checkmark{} / \checkmark{} \\
        \rowcolor{gray!10} 
        \textit{attacker\_addresses} & Wallet addresses controlled by attackers & 
        \href{https://etherscan.io/address/0x47666fab8bd0ac7003bce3f5c3585383f09486e2}{\textit{0x4766}} & \checkmark{} / \checkmark{} \\
        \textit{victim\_addresses} & Wallet addresses of victims or source accounts & \href{https://etherscan.io/address/0x1Db92e2EeBC8E0c075a02BeA49a2935BcD2dFCF4}{\textit{0x1Db9}}& \checkmark{} / \checkmark{} \\
        \rowcolor{gray!10} 
        \textit{stolen\_usd} & Estimated value of stolen funds in USD & \$1{,}500{,}000{,}000 & \checkmark{} / \checkmark{} \\
        \textit{stolen\_token} & Types and amounts of stolen tokens & "ETH": 401{,}000,
        "mETH": 8{,}000, 
        "cmETH": 15{,}000,
        "stETH": 90{,}000 & \checkmark{} / \checkmark{} \\
        \bottomrule
      \end{tabular}
  }
  \vspace{2pt}
  \footnotesize
  \textit{Note.} Each check mark denotes agreement with one human annotator.
\end{table}

\begin{figure}[t]
    \centering
    \includegraphics[width=\columnwidth]{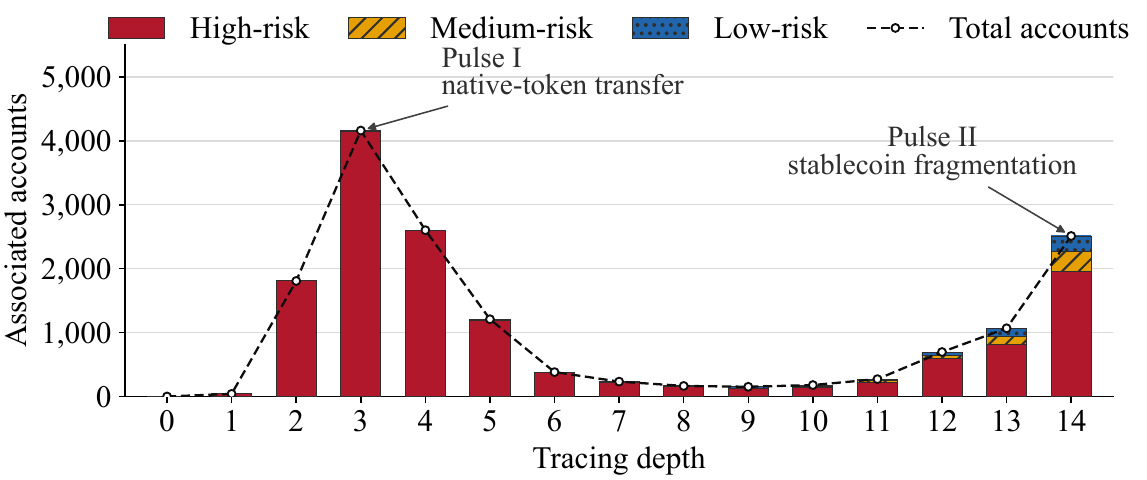}
\caption{\textcolor{black}{Layer-wise account distribution in the Bybit Hack. The bars show risk composition by tracing depth, while the dashed curve shows the total number of associated accounts. The two peaks correspond to early native-token movement and later stablecoin fragmentation.}}
    \label{fig:bybit_bimodal_pulsing}
\end{figure}

\begin{table*}[t]
\centering
\caption{Comprehensive comparison with tracing baselines across five incidents. Bold indicates the best value and underline indicates the second-best value.}
\label{tab:baseline_summary}
\scriptsize
\setlength{\tabcolsep}{2.1pt}
\renewcommand{\arraystretch}{1.4}
\resizebox{\textwidth}{!}{%
\rowcolors{2}{tablerowgray}{white}
\begin{tabular}{l*{20}{c}}
\toprule
\textbf{Method} &
\multicolumn{4}{c}{\textbf{Bybit}} &
\multicolumn{4}{c}{\textbf{Li.Fi}} &
\multicolumn{4}{c}{\textbf{HTX \& Heco}} &
\multicolumn{4}{c}{\textbf{Ronin}} &
\multicolumn{4}{c}{\textbf{Vulcan Forged}} \\
\cmidrule(lr){2-5}
\cmidrule(lr){6-9}
\cmidrule(lr){10-13}
\cmidrule(lr){14-17}
\cmidrule(lr){18-21}
& $R_a$ & $P_a$ & F1 & $R_f$
& $R_a$ & $P_a$ & F1 & $R_f$
& $R_a$ & $P_a$ & F1 & $R_f$
& $R_a$ & $P_a$ & F1 & $R_f$
& $R_a$ & $P_a$ & F1 & $R_f$ \\
\midrule
Tiles
& 1.30 & \textbf{100.00} & 2.56 & 0.00
& 0.83 & \textbf{100.00} & 1.65 & 0.00
& 9.52 & \textbf{100.00} & 17.39 & 0.00
& 11.11 & \textbf{100.00} & 20.00 & 0.00
& \underline{50.00} & \textbf{100.00} & 66.67 & 0.00 \\
XBlockFlow
& 3.90 & 40.68 & 7.12 & 5.34
& \underline{69.42} & 5.97 & 10.99 & \underline{87.78}
& \underline{95.24} & \underline{58.82} & \underline{72.73} & \underline{62.90}
& 33.33 & 0.39 & 0.36 & 4.94
& \textbf{100.00} & \underline{76.47} & \underline{86.67} & 1.22 \\
Poison
& \underline{70.13} & 57.14 & \underline{62.97} & \underline{21.53}
& 37.19 & 11.60 & 17.70 & 41.21
& 61.90 & 50.00 & 55.32 & 48.42
& 55.56 & 38.10 & \underline{45.20} & 29.52
& \textbf{100.00} & 52.63 & 68.96 & \underline{26.24} \\
Haircut
& 57.14 & 30.00 & 39.34 & 17.70
& 34.71 & \underline{50.60} & \underline{41.18} & 59.76
& 19.05 & \textbf{100.00} & 32.00 & 34.64
& \underline{66.67} & 15.00 & 24.49 & 5.27
& \textbf{100.00} & 75.00 & 85.71 & 17.93 \\
TPP
& 59.74 & 38.10 & 46.53 & 12.37
& \underline{69.42} & 18.71 & 29.49 & 45.84
& 66.67 & 13.33 & 22.22 & 11.12
& 11.11 & 28.57 & 16.00 & \underline{35.47}
& \textbf{100.00} & 30.00 & 46.15 & 4.71 \\
APPR
& 55.84 & 23.81 & 33.39 & 1.37
& 26.45 & 12.60 & 17.07 & 48.40
& \textbf{100.00} & 54.17 & 70.27 & 24.91
& 33.33 & 0.00 & 0.00 & 0.00
& \textbf{100.00} & 60.42 & 75.33 & 16.39 \\
\textsc{RiskTagger}
& \textbf{97.33} & \underline{98.69} & \textbf{98.01} & \textbf{89.07}
& \textbf{100.00} & \textbf{100.00} & \textbf{100.00} & \textbf{96.66}
& \underline{95.24} & \textbf{100.00} & \textbf{97.56} & \textbf{99.10}
& \textbf{100.00} & \underline{91.27} & \textbf{95.44} & \textbf{95.77}
& \textbf{100.00} & \textbf{100.00} & \textbf{100.00} & \textbf{85.62} \\
\bottomrule
\end{tabular}}
\rowcolors{2}{}{}
\vspace{0.5mm}
\begin{minipage}{0.96\textwidth}
\footnotesize
\end{minipage}
\end{table*}

\subsection{Metrics}
Let $\mathcal{A}_{GT}$ denote the public GT account set, and let $\mathcal{A}_{+}$ denote accounts labeled \textit{high} or \textit{medium} risk by \textsc{RiskTagger}. Unless otherwise noted, \textit{high}/\textit{medium} are treated as positive laundering-related outputs, while \textit{low}/\textit{none} are treated as negative outputs. Expert review is used for precision, false-positive analysis, and validation of \textsc{RiskTagger} outputs not covered by public labels. The unified criteria for expert review are provided in the Supplementary Material. The metrics are grouped by module as follows.

\subsubsection{Extractor metrics} The key-clue extractor is evaluated by entity precision and information coverage against a manually checked gold record. Entity precision measures the fraction of extracted entities that are correct. Entity precision is denoted by $P_e$. Information coverage is defined as \[ R_{cov}=\frac{E_{full}+0.5E_{part}}{E_{all}}, \] where $E_{full}$, $E_{part}$, and $E_{all}$ denote fully covered, partially covered, and all required case fields, respectively. 

\subsubsection{Tracer metrics} The tracer is evaluated by the following metrics. Maximum tracing depth records the largest recursive depth reached by \textsc{RiskTagger}. The number of high/medium-risk accounts is $|\mathcal{A}_{+}|$. Address recall is defined as \[ R_a=\frac{|\mathcal{A}_{+}\cap\mathcal{A}_{GT}|}{|\mathcal{A}_{GT}|}. \] Fund recall is defined as $ R_f= F_{rec} / F_{total}, $ where $F_{total}$ is the case-specific fund denominator used in each experiment. Address precision, false positive rate (FPR), and F1-score are computed on expert-reviewed samples.
FPR is omitted from the tracing-baseline comparison because tracing baselines only output suspected laundering accounts, making true negatives undefined.
\textit{New $\mathcal{A}_{+}$ accounts} denote high/medium-risk accounts that are not already included in the public GT set. We further report Krippendorff's $\alpha$ to measure the agreement between \textsc{RiskTagger} labels and expert-reviewed labels beyond chance: $ \alpha = 1 - \frac{D_o}{D_e}, $ where $D_o$ is the observed disagreement and $D_e$ is the expected disagreement under random labeling. A larger $\alpha$ indicates stronger label agreement.


\subsubsection{Explainer metrics} The evidence explainer is evaluated by mean 1--5 scores on Accuracy, Logic, Professionalism, and Actionability. We also report an Overall score computed as the average over these four dimensions. Report quality is assessed through a multi-agent cross-scoring protocol, with the scoring prompt and rubric provided in the supplement.

\subsection{Implementation.}
\textsc{RiskTagger} is implemented in Python and orchestrates the three LLM-involved components through a LangChain-based workflow~\textcolor{black}{\cite{langchainDocs2026}}. 
The default LLM backend is Qwen3-Max for the key-clue extractor, laundering tracer, and evidence explainer~\textcolor{black}{\cite{yang2025qwen3technicalreport,alibabaModelStudioModels2026}}. We use it because it provides long-context support and stable structured reasoning over recursive transaction evidence; the evaluation does not claim that this model is globally superior. The memory module uses text-embedding-v2 embeddings and ChromaDB for vector retrieval~\textcolor{black}{\cite{alibabaEmbeddingDocs2026,chromaDocs2026}}. Runtime is reported as wall-clock time under 16 parallel workers; the current experiment notes record an average of 1.015 seconds per address.
%
We keep the implementation settings fixed across cases unless noted. The LLM temperature is 0.3. The expansion controller removes micro-transfer noise with a dust threshold $\tau$ of \textcolor{black}{0.05 ETH or equivalent}, caps fanout at 100 addresses per tracing depth, keeps receivers covering 99\% of outgoing value ($\theta$), and uses a 50-day temporal window ($T$). Recursive tracing uses a default maximum tracing depth $D=15$ and permits $D=20$ for extreme cases. Memory retrieval uses $m=1$ to retrieve the most relevant prior decision context. These settings are reported to make the evaluation reproducible; they are not claimed as globally optimal parameters.

\section{Experiment Results and Insights} \label{sec:experiment_results}

\subsection{RQ1: Main Case -- Bybit Hack}
\label{subsec:result_bybit}

The Bybit Hack is used as the primary incident because it stresses the full \textsc{RiskTagger} workflow: clue extraction from public materials, recursive tracing over a large deep laundering graph, and evidence organization for analyst review. The key challenge is not only to recover more accounts, but also to maintain forensic precision when the trace expands across services, dust transfers, and high-degree activity.

\noindent\textbf{Clue recovery.}
The extractor  is evaluated against a manually checked gold record constructed by independent reviewers with arbitration. As shown in Table~\ref{tab:rq1-case-example}, \textsc{RiskTagger} recovers the main case fields, including seed addresses, affected assets, attack context, and initial fund-flow clues. Accurate clue recovery is therefore important because early extraction errors can propagate through subsequent recursive expansion.

\noindent\textbf{Tracing reliability.} 
Starting from the seed addresses, the tracer constructs a risk-labeled graph with maximum tracing depth 15 and 15,497 associated accounts. 
Among them, 14,952 accounts are labeled as high/medium risk, while 545 accounts are labeled as low/no risk. A 5\% sample is drawn across all risk levels, providing a statistically guaranteed margin of error within $\pm 3.38\%$ at the 95\% confidence level.
As reported in Table~\ref{tab:bybit_performance}, \textsc{RiskTagger} achieves 97.33\% address recall, recovers 89.07\% of the ETH-denominated evaluation denominator, and obtains 98.69\% sampled precision with an F1-score of 98.01\%. The expert agreement reaches a Krippendorff's $\alpha$ of 0.7848, indicating substantial consistency between system labels and independent expert judgments. These results suggest that the controlled expansion strategy can recover most known laundering accounts and funds without collapsing into an overly broad graph search.

\noindent\textbf{Evidence insight.}
Beyond account-level recovery, the explainer converts the traced graph into a structured forensic report and exposes a topology-level laundering pattern. 
As shown in Fig.~\ref{fig:bybit_bimodal_pulsing}, the trace exhibits a bimodal pulsing structure. An early peak dominated by high-risk native-token movement, followed by a low-activity interval and a deeper peak with more mixed risk levels. This pattern is consistent with a transition from rapid initial fund movement to later stablecoin fragmentation and downstream dispersion. The insight is important because it links the quantitative recovery results to an analyst-reviewable laundering structure, rather than leaving the output as a large suspicious-address list.

\begin{insightbox}
    \textit{\textbf{Answer to RQ1:} The Bybit case shows that \textsc{RiskTagger} can maintain high forensic coverage in deep laundering scenarios while preserving high sampled precision through evidence constraints and controlled expansion.}
\end{insightbox}

\subsection{RQ2: Generalization Across Additional Cases}
\label{subsec:generalization}

We further evaluate whether \textsc{RiskTagger} generalizes beyond the Bybit Hack to four additional incidents. These cases are representative because they cover common but distinct entry points and fund-transfer structures in post-incident Web3 laundering forensics, making them a stress test for whether the system depends on a single incident shape. 
%
Table~\ref{tab:generalization_metrics} reports the results at the levels of key-clue extraction, laundering tracing, and evidence explanation. 



\begin{table}[t]
\centering
\caption{Generalization performance across four additional incidents.}
\label{tab:generalization_metrics}
\footnotesize
\setlength{\tabcolsep}{2pt}
\renewcommand{\arraystretch}{1.3}
\begin{tabular*}{\columnwidth}{@{\extracolsep{\fill}}lcccc@{}}
\toprule
\textbf{Metric} & \textbf{Vulcan} & \textbf{Ronin} &
\textbf{HTX\&Heco} & \textbf{Li.Fi} \\
\midrule
\multicolumn{5}{l}{\textbf{I. Key-clue Extractor}} \\
$P_e$ & 100.00\% & 100.00\% & 91.00\% & 100.00\% \\
$R_{cov}$ & 100.00\% & 100.00\% & 100.00\% & 100.00\% \\
\midrule
\multicolumn{5}{l}{\textbf{II. Laundering Tracer}} \\
Max depth & 3 & 15 & 4 & 19 \\
$|\text{New}~\mathcal{A}_{+}|$ & 20 & 884 & 13 & 103 \\
$R_f$ & 85.62\% & 95.77\% & 99.10\% & 96.66\% \\
$R_a$ & 100.00\% & 100.00\% & 95.24\% & 100.00\% \\
$P_a$ & 100.00\% & 91.27\% & 100.00\% & 100.00\% \\
F1 & 100.00\% & 95.44\% & 97.56\% & 100.00\% \\
\midrule
\multicolumn{5}{l}{\textbf{III. Evidence Explainer}} \\
Accuracy & 4.14/5 & 3.57/5 & 4.57/5 & 3.71/5 \\
Logic & 4.86/5 & 4.43/5 & 4.86/5 & 4.71/5 \\
Professionalism & 4.86/5 & 4.71/5 & 4.86/5 & 4.57/5 \\
Actionability & 4.71/5 & 4.43/5 & 4.57/5 & 4.43/5 \\
Overall & 4.43/5 & 4.21/5 & 4.57/5 & 4.21/5 \\
\bottomrule
\end{tabular*}
\vspace{0.5mm}
\begin{minipage}{0.96\columnwidth}
\footnotesize
\end{minipage}
\end{table}


\noindent\textbf{Vulcan Forged incident (2021).}
The Vulcan Forged incident represents a short-horizon cash-out case in which laundering complexity mainly arises from fund fragmentation rather than path length. \textsc{RiskTagger} identifies 22 high-risk accounts, including 20 accounts beyond the public ground-truth set. Although the reconstructed trace terminates within tracing depth 3, the evidence explainer further shows that 2,000,000 stolen PYR tokens were split into 509 small-value transfers before being deposited into centralized exchanges, thereby weakening the linkage between downstream accounts and the original attack incident. 
This case shows that tracing should preserve the incident context of fund movements. Low-value transfers may still be forensic-relevant when they are temporally close to the incident and semantically linked to the stolen assets.

\begin{figure}[t]
    \centering
    \includegraphics[width=\columnwidth]{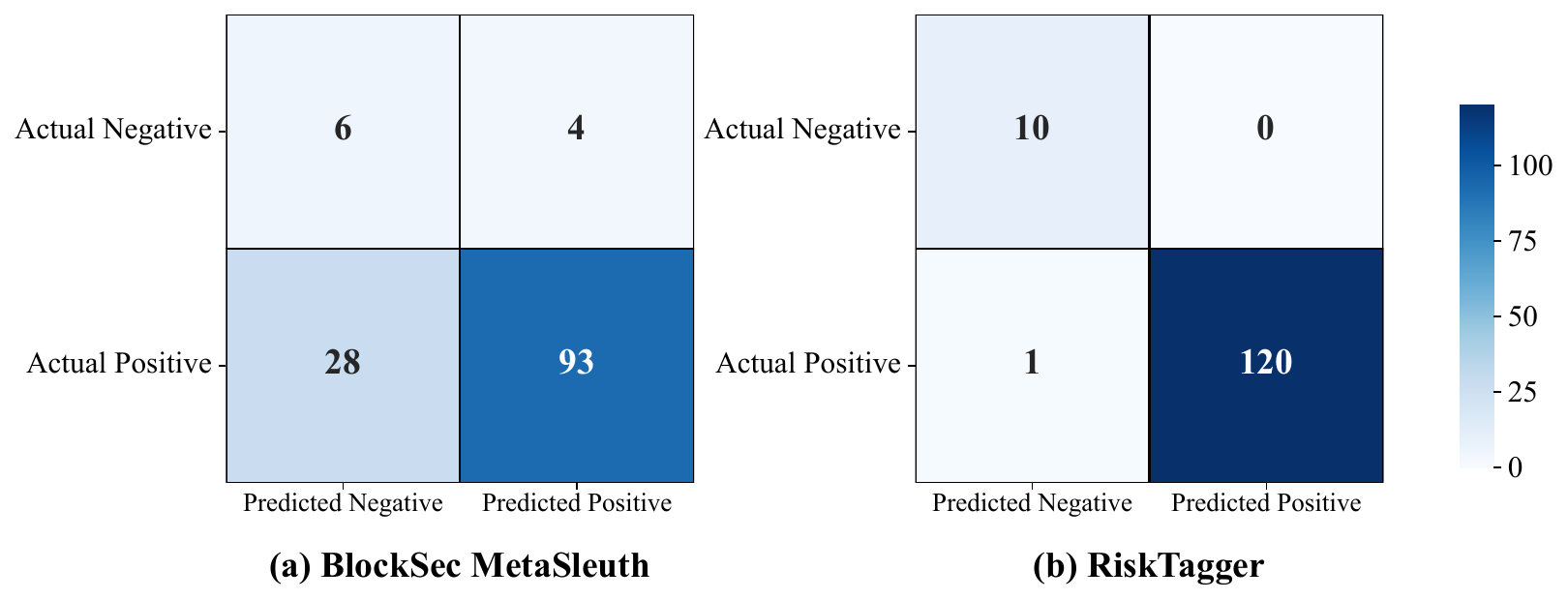}
\caption{Comparison on the laundering tracing results of Li.Fi incident. }
    \label{fig:lifi_disagreement}
\end{figure}

\noindent\textbf{Ronin Network incident (2022).}
Ronin exhibits a deeper and denser bridge-laundering pattern, where the fund movement extends substantially beyond short cash-out traces. \textsc{RiskTagger} reaches tracing depth 15 and identifies 884 high-risk accounts that are not covered by the public ground-truth set. The Memory component further helps preserve cross-case consistency by retrieving historical risk labels and AML-oriented judgments. Based on this mechanism, \textsc{RiskTagger} finds 19 overlapping accounts between the Ronin and Bybit traces, indicating possible reuse of downstream laundering accounts or service infrastructure. 
This finding is consistent with the U.S. Treasury disclosure that Blender.io was used in relation to the Ronin/Axie Infinity stolen funds~\cite{treasury2022blenderSanctions}. Therefore, the overlap provides useful forensic evidence of infrastructure reuse, rather than a standalone attribution conclusion.

\noindent\textbf{HTX \& Heco incident (2023).}
HTX \& Heco incident highlights the difficulty of tracing laundering flows in complex DeFi interactions, where legitimate routers, high-frequency contracts, and attack-related transfers can be closely connected. \textsc{RiskTagger} expands the trace to tracing depth 4, identifies 13 new risk accounts, and recovers 99.10\% of the stolen funds. 
This result is supported by the expansion controller, which filters candidate branches using transaction-level constraints and fund-dominance selection before further expansion. The case shows that effective tracing in such settings requires both sufficient coverage and controlled frontier selection, rather than unbounded graph expansion.

\noindent\textbf{Li.Fi incident (2024).}
Li.Fi incident is retained as the main disagreement-analysis case because public results from BlockSec MetaSleuth report~\cite{metasleuth2024lifi} are available for comparison. As shown in Fig.~\ref{fig:lifi_disagreement}, among 131 accounts, MetaSleuth misclassifies 28 suspicious accounts as low-risk accounts and 4 low-risk accounts as high-risk accounts; in contrast, \textsc{RiskTagger} reduces the disagreement to only one false negative and achieves zero false positives. In this case, \textsc{RiskTagger} reaches tracing depth 19 and identifies 103 risk accounts beyond the public ground-truth set. Its only missed high-risk account is an eXch hot wallet, whose long activity history, large transaction volume, and highly diverse counterparties make it resemble a mature service account. In addition, \textsc{RiskTagger} reveals a deterministic denomination-splitting pattern, in which the stolen funds are partitioned into 100, 10, 1, and 0.1 ETH units before entering Tornado Cash. 
This case provides a diagnostic view of both the strengths and limitations of \textsc{RiskTagger} under service-heavy laundering paths.

\begin{table}[t]
\centering
\caption{Component ablation of \textsc{RiskTagger} on the Li.Fi incident.}
\label{tab:ablation_components}
\footnotesize
\setlength{\tabcolsep}{3.0pt}
\renewcommand{\arraystretch}{1.3}
\rowcolors{2}{tablerowgray}{white}
\begin{tabular}{p{0.30\columnwidth}p{0.14\columnwidth}p{0.14\columnwidth}p{0.11\columnwidth}p{0.13\columnwidth}}
\hline
\textbf{Variant} & \textbf{$R_a$} & \textbf{$P_a$} & \textbf{FPR} & \textbf{$R_f$} \\
\hline
w/o Controller & 61.16\% & 8.62\% & 93.49\% & 8.33\% \\
w/o Reflection & 74.38\% & 97.83\% & 20.00\% & 94.57\% \\
w/o Memory & 89.26\% & 98.18\% & 20.00\% & 94.90\% \\
w/o CoT & 85.95\% & 99.05\% & 10.00\% & 95.74\% \\
\textbf{Full system} & \textbf{99.17\%} & \textbf{100.00\%} &
\textbf{0.00\%} & \textbf{96.66\%} \\
\hline
\end{tabular}
\rowcolors{2}{}{}
\end{table}

\begin{insightbox}
    \textit{\textbf{Answer to RQ2:} 
    The results indicate that \textsc{RiskTagger} generalizes beyond the Bybit Hack incident to laundering behaviors with different tracing depths, assets, protocols, and service boundaries, while keeping false positives low under different complexity levels.
    }
\end{insightbox}

\begin{table*}[t]
\centering
\caption{LLM backend comparison of \textsc{RiskTagger} on the Li.Fi incident.}
\label{tab:llm_comparison}
\scriptsize
\setlength{\tabcolsep}{3.2pt}
\renewcommand{\arraystretch}{1.3}
\rowcolors{2}{tablerowgray}{white}
\begin{tabular}{p{0.19\textwidth}p{0.09\textwidth}p{0.10\textwidth}p{0.09\textwidth}p{0.10\textwidth}p{0.10\textwidth}p{0.09\textwidth}p{0.10\textwidth}}
\hline
\textbf{LLM} & \textbf{Time (s)} & \textbf{Tokens (K)} &
\textbf{Cost (\$)} & \textbf{$P_a$} & \textbf{$R_a$} &
\textbf{FPR} & \textbf{$R_f$} \\
\hline
DeepSeek V4 Flash & 522 & 381.40 & 0.078 & 99.17\% & 98.35\% & 10.00\% & 95.86\% \\
DeepSeek V4 Pro & 1810 & 479.10 & 0.317 & 99.14\% & 95.04\% & 10.00\% & 95.83\% \\
GLM 5.1 & 889 & 462.15 & 1.031 & 99.17\% & 99.17\% & 10.00\% & 95.86\% \\
Seed-OSS 36B Instruct & 997 & 576.71 & 0.237 & 99.17\% & 98.35\% & 10.00\% & 95.86\% \\
Claude Opus 4.7 & 506 & 214.93 & 0.519 & 100.00\% & 87.60\% & 0.00\% & 96.66\% \\
GPT 5.4 & 539 & 300.74 & 1.410 & 99.14\% & 95.04\% & 10.00\% & 95.83\% \\
Gemini 3.1 Pro Preview & 533 & 493.89 & 1.407 & 99.15\% & 96.69\% & 10.00\% & 95.84\% \\
\textbf{Qwen3-Max} & \textbf{372} & 315.20 & 0.171 &
\textbf{100.00\%} & \textbf{99.17\%} & \textbf{0.00\%} &
\textbf{96.66\%} \\
\hline
\end{tabular}
\rowcolors{2}{}{}
\vspace{0.5mm}
\begin{minipage}{0.96\textwidth}
\footnotesize
\end{minipage}
\end{table*}

\subsection{RQ3: Comparison with Tracing Baselines}
\label{subsec:baseline_comparison}

We compare the tracing output of \textsc{RiskTagger} with three groups of representative tracing baselines:
\begin{itemize}
    \item dynamic community detection: Tiles~\cite{Tiles};
    \item taint-propagation methods: XBlockFlow~\cite{wu2023toward}, Poison~\cite{Poison&Haircut}, Haircut~\cite{Poison&Haircut}, and TPP~\cite{TPP};
    \item personalized ranking: APPR~\cite{APPR}.
\end{itemize}
All methods use the same case seeds and transaction scope. 

Table~\ref{tab:baseline_summary} summarizes the results over five real incidents. \textsc{RiskTagger} achieves 93.24\% average fund recall, improving over the best incident-level baseline by 8.88--67.54 percentage points. 
The baselines reveal complementary weaknesses.
Tiles returns a small set of high-confidence accounts,
which leads to high sampled precision, but its address recall and fund recall are limited in most cases. Taint-propagation methods can expand to more downstream accounts and improve recall under certain settings, yet they are sensitive to dust transfers, fan-out behavior, and service-like accounts. For example, XBlockFlow achieves 69.42\% address recall and 87.78\% fund recall on Li.Fi, while its sampled precision is 5.97\%. Poison and TPP recover more accounts than Tiles in several cases, but their F1-scores and fund recall differ substantially across incidents. APPR mainly relies on local graph proximity and also shows case-dependent behavior. It obtains 100.00\% address recall on HTX \& Heco, but recovers 24.91\% of the funds and achieves limited performance on Ronin.



By conditioning expansion on case context, transaction semantics, memory, and controller, \textsc{RiskTagger} maintains a more stable trade-off between address-level identification and fund-flow recovery. The comparison suggests that topology-only ranking and fixed taint-propagation rules are insufficient for heterogeneous laundering traces.

\begin{insightbox}
    \textit{\textbf{Answer to RQ3:} \textsc{RiskTagger} outperforms tracing baselines because it controls where to expand, not merely how far to expand. This enables more stable recovery of laundering-related accounts and funds under heterogeneous Web3 forensic scenarios.
    }
\end{insightbox}

\subsection{RQ4: Component Contribution and Backend Sensitivity}

\subsubsection{Component Ablation}
\label{subsec:component_ablation}

We use Li.Fi incident as the ablation case because its expert-reviewed labels contain both laundering and benign service-like accounts, making it suitable for testing boundary decisions. 

Table~\ref{tab:ablation_components} reports the component-level results. The largest degradation occurs after removing the controller, because recursive tracing loses its operational boundary and admits low-value, weakly related, or cyclic branches. This result is consistent with the design of the controller, which preserves dominant fund flows while suppressing noisy expansion. The other components show complementary effects. Removing Reflection mainly reduces recall, indicating that post-decision consistency checking helps recover from unstable intermediate reasoning. Removing Memory increases false positives because the reasoner loses historical case context and recurring infrastructure cues. Removing CoT also weakens recall, suggesting that multi-signal laundering behaviors require explicit evidence-by-evidence reasoning rather than one-shot labeling.

\subsubsection{LLM Backend Comparison}
\label{subsec:llm_backend_comparison}

We also compare LLM backends on the Li.Fi setup. This experiment is used to
select the default backend for \textsc{RiskTagger}; it is not a claim about global model
superiority. The same prompts, runtime settings, and evaluation labels are used across
models. Table~\ref{tab:llm_comparison} compares the tested LLM backends under the same Li.Fi setup. The \textit{Time (s)} column reports LLM-only wall-clock inference time
under the same 16-worker parallel execution setting. It excludes blockchain data
fetching, graph construction, manual review, and post-hoc validation, and thus
does not represent end-to-end forensic processing time.

Qwen3-Max gives the best balance in this setup: it preserves zero sampled false
positives, matches the strongest precision and recall profile, and has the
shortest runtime among the tested backends. Some models achieve competitive
recall but introduce false positives, while Claude Opus 4.7 is conservative
and loses recall. We therefore use Qwen3-Max as the default backend in the
main experiments, while treating the architecture itself as backend-agnostic.

\begin{insightbox}
    \textit{\textbf{Answer to RQ4:} The ablation results support the design choice of \textsc{RiskTagger}: The memory-augmented reasoner assigns evidence-grounded risk labels, the reflection design stabilizes the decision, and the expansion controller keeps recursive tracing bounded.}
\end{insightbox}


\section{Discussion}
\label{sec:discussion}

\noindent\textbf{The scope of our study.}
\textsc{RiskTagger} is designed for post-incident forensic triage under a
public-evidence setting. Its outputs should be interpreted as
evidence-organized leads for analyst review, not as final attribution, legal
judgment, or proof of real-world identity. This boundary is important because
the system observes public incident materials and public on-chain records, but
does not assume privileged exchange records, private investigative data,
personally identifiable information, or off-chain account ownership.


\noindent\textbf{Dependency on external tools.}
\textsc{RiskTagger} depends on the availability and quality of public on-chain access. The current implementation focuses on EVM-compatible chains and cross-chain continuations that can be supported by transaction-association tools. When a bridge, service, or chain does not expose sufficient public evidence, the system records a boundary instead of inferring an unsupported continuation. 

\noindent\textbf{Limitations.}
The main limitation comes from the public nature of the evidence. Incident
reports and blockchain-explorer labels can be incomplete, delayed, or
inconsistent across sources. \textsc{RiskTagger} therefore uses public labels for
case-level recall and expert review for outputs not covered by public labels,
but unlabeled downstream accounts should still be treated as leads rather than
confirmed laundering entities. The same principle applies to Evidence
Explainer outputs: the report organizes paths, labels, and representative
evidence, while final judgment remains with human analysts.

\noindent\textbf{Robustness discussion.}
RiskTagger balances recall and noise control in graph expansion through parameters such as tracing depth and dust thresholds, but the system does not rely on parameter secrecy. Even if attackers attempt to evade tracing by splitting funds, delaying transfers, increasing fan-out, or routing through service accounts, on-chain evidence cannot be deleted and may instead expose new anomalous patterns. Therefore, the system performs joint reasoning over asset matching, temporal feasibility, contract routing, fund-dominance relationships, and loop checks. These parameters are treated as a tunable recall–noise budget; paths with insufficient evidence are marked only as evidence boundaries rather than confirmed laundering paths.


\section{Conclusion}
\label{sec:conclusion}

This paper studies post-incident Web3 money-laundering forensics and presents \textsc{RiskTagger}, an LLM-guided forensic framework. \textsc{RiskTagger} constrains the LLM within a closed-loop forensic workflow of clue extraction, evidence-based tracing, and report generation, thereby automatically recovering case-related risk accounts and reviewable explanations. In the Bybit hack main incident, \textsc{RiskTagger} traces up to 15 hops, constructs a risk-labeled graph with 15,497 associated accounts, and achieves 97.33\% address recall, 89.07\% fund recall, and 98.69\% sampled precision. The cross-case study further shows that \textsc{RiskTagger} adapts to real-world Web3 security incidents across different years, attack vectors, and laundering structures, indicating that it is not effective only for a single case or path pattern. Importantly, \textsc{RiskTagger} is not intended to replace human investigators or provide standalone attribution conclusions; instead, it offers structured, evidence-supported forensic assistance for analyst review.

\bibliographystyle{IEEEtran}
\bibliography{refs}

\begin{IEEEbiography}[{\includegraphics[width=1in,height=1.25in,clip,keepaspectratio]{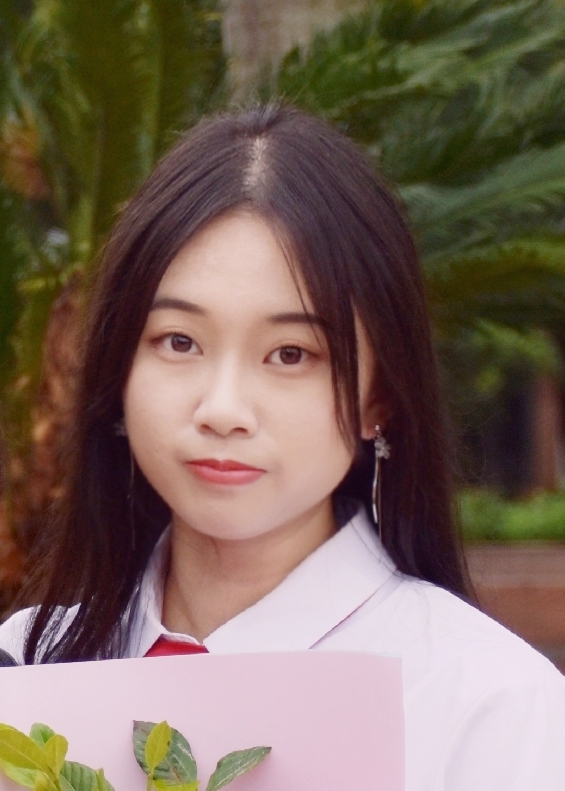}}]{Dan Lin} (Member, IEEE)
received the Ph.D. degree from Sun Yat-sen University in 2024. She is currently an Assistant Researcher and Post-Doctoral Fellow with the School of Software Engineering, Sun Yat-sen University. Her research interests include blockchain security, cross-chain bridge analysis, cryptocurrency anti-money laundering, and network science.
Dr. Lin also serves as an Executive Committee Member of the Technical Committee on Network and Information Security, and the Technical Committee on Blockchain of the China Computer Federation (CCF). She has served as a Program Chair for the ISSTA 2025 Workshop and KSEM 2025, and as a Distinguished Session Chair of the International Conference on Intelligence (ICI 2026). Her work has received several awards, including the IEEE ISCAS 2024 Best Paper Award.
\end{IEEEbiography}

\vspace{-7ex}

\begin{IEEEbiography}[{\includegraphics[width=1in,height=1.25in,clip,keepaspectratio]{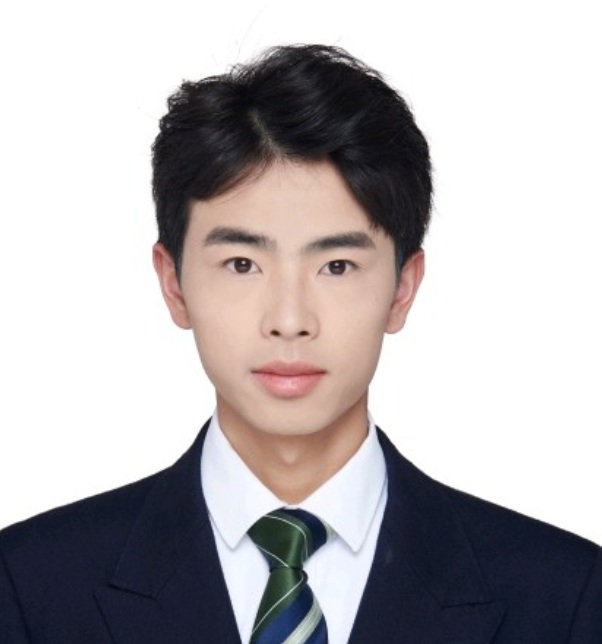}}]{Yanli Ding}
received the B.Eng. degree in Computer Science and Technology from Xiamen
University, Xiamen, China, in 2025. He is currently pursuing the master's
degree with the School of Software Engineering, Sun Yat-sen University,
Guangzhou, China. His current research interests include blockchain and
anti-money laundering.
\end{IEEEbiography}

\vspace{-8ex}

\begin{IEEEbiography}[{\includegraphics[width=1in,height=1.25in,clip,keepaspectratio]{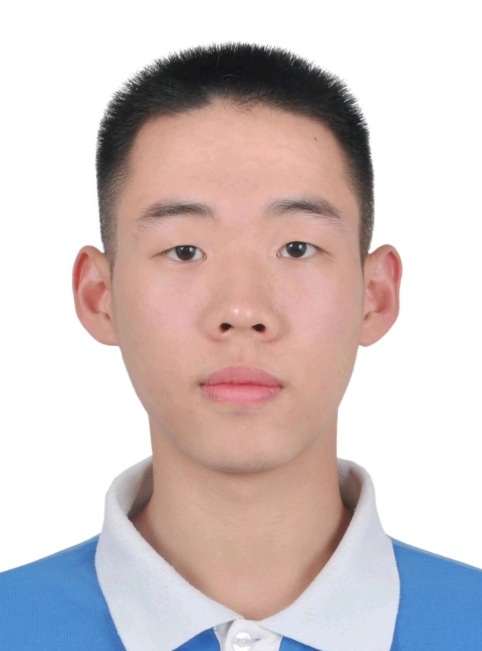}}]{Weipeng Zou}
received the B.Eng. degree in Software Engineering from Sun Yat-sen
University, Guangzhou, China, in 2026. He will pursue his master's degree with
the School of Software Engineering, Sun Yat-sen University, Guangzhou, China.
His research interests include blockchain technology, anti-money laundering,
and stablecoins.
\end{IEEEbiography}

\vspace{-8ex}

\begin{IEEEbiography}[{\includegraphics[width=1in,height=1.25in,clip,keepaspectratio]{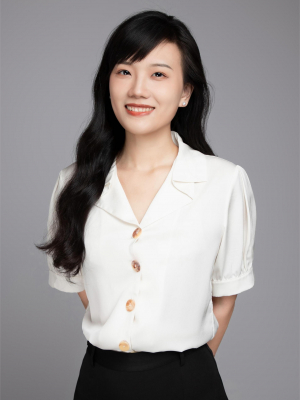}}]{Jiajing Wu}
(Senior Member, IEEE) received the B.Eng. degree in communication engineering from Beijing Jiaotong University, Beijing, China, in 2010, and the Ph.D. degree from Hong Kong Polytechnic University, Hong Kong, in 2014. She was awarded the Hong Kong Ph.D. Fellowship Scheme during her Ph.D. study in Hong Kong (2010--2014). She is currently a Professor with the Shenzhen Research Institute, Sun Yat-sen University, Shenzhen, China. Her research focus includes blockchain, graph mining, and network science. She serves as an Associate Editor for {\sc IEEE Network.}
\end{IEEEbiography}

\vspace{-8ex}

\begin{IEEEbiography}[{\includegraphics[width=1in,height=1.25in,clip,keepaspectratio]{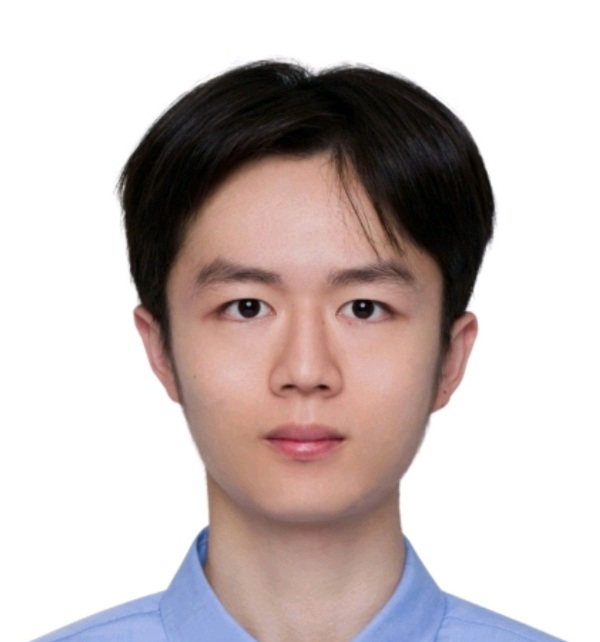}}]{Zhiying Wu} received the B.Eng. degree in computer science and technology from Wuhan University of Technology, Wuhan, China, in 2021. He received the Ph.D. degree with the School of Computer Science and Engineering, Sun Yat-sen University, in 2026. His current research interests include blockchain, network science, and machine learning with graphs.
\end{IEEEbiography}

\vspace{-6ex}

\begin{IEEEbiography}[{\includegraphics[width=1in,height=1.25in,clip,keepaspectratio]{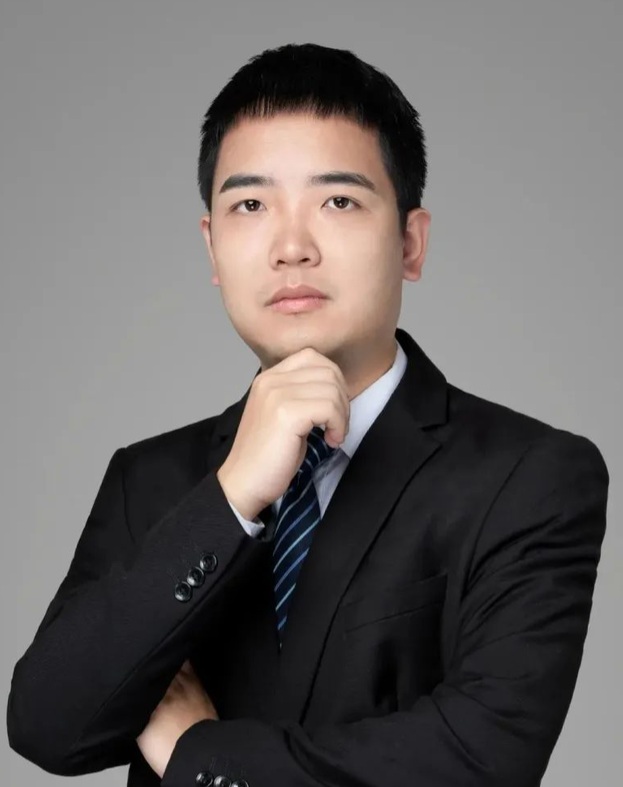}}]{
Jiachi Chen} is currently a ZJU 100 Young Professor in the College of Computer Science and Technology at Zhejiang University. Before joining Zhejiang University, he served as an Associate (tenured) Professor in the School of Software Engineering at Sun Yat-senUniversity. His research interests focus on Web3 security, AI for Web3, and software testing.

\end{IEEEbiography}

\vspace{-5ex}

\begin{IEEEbiography}[{\includegraphics[width=1in,height=1.25in,clip,keepaspectratio]{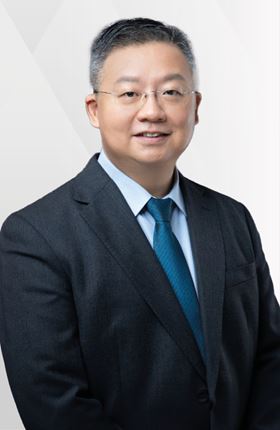}}]{Xiapu Luo} is a professor with the Department of Computing, The Hong Kong Polytechnic University. His research focuses on blockchain and smart contracts security, mobile and IoT security, network security and privacy, and software engineering withpapers published in top-tier security, software engineering, and networking venues. His research led to more than ten best/distinguished paper awards, including ACM CCS'24 Distinguished Paper Award,three ACM SIGSOFT Distinguished Paper Awardsin ICSE'24, ISSTA'22, and ICSE'21, Best DeFi Papers Award 2023, Best Paper Award in INFOCOM' 18, Best Research Paper Award in ISSRE' 16, etc., and several awards from the industry. He received the BOCHK Science and Technology Innovation Prize (FinTech) for his contribution to blockchain security. He regularly serves in the program committees of top security and software engineering conferences and received the Top Reviewer Award from CCS'22 and the Distinguished TPC Member Award from INFOCOM'23 and INFOCOM'24.
\end{IEEEbiography}

\vspace{-6ex}

\begin{IEEEbiography}[{\includegraphics[width=1in,height=1.25in,clip,keepaspectratio]{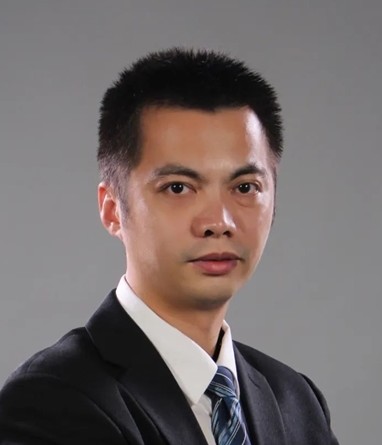}}]{Zibin Zheng} (Fellow, IEEE) is currently a Professor and the Dean with the School of Software Engineering, Sun Yat-sen University, Guangzhou, China. He authored or coauthored more than 200 international journal and conference papers, including one ESI hot paper and ten ESI highly cited papers. According to Google Scholar, his papers have more than 28,000 citations. His research interests include blockchain, software engineering, and services computing. He was the BlockSys’19 and CollaborateCom16 General Co-Chair, SC2’19, ICIOT18 and IoV14 PC Co-Chair. He is a Fellow of the IET. He received several awards, including the Top 50 Influential Papers in Blockchain of 2018, the ACM SIGSOFT Distinguished Paper Award at ICSE2010, the Best Student Paper Award at ICWS2010.
\end{IEEEbiography}


\end{document}